\documentclass[journal,10pt]{IEEEtran}

\ifCLASSINFOpdf

\else

\fi

\usepackage[utf8]{inputenc}
\usepackage{url}
\usepackage{cite}
\usepackage{hyperref}
\usepackage{graphicx}
\usepackage[nodisplayskipstretch]{setspace}
\usepackage{amsfonts} 
\usepackage{lipsum}
\usepackage{amsmath}
\usepackage{dsfont}
\usepackage{amssymb}
\usepackage{bbm}
\usepackage{amsmath}

\DeclareMathOperator*{\limisup}{lim\,sup}
\DeclareMathOperator*{\amin}{\mbox{min.}}
\usepackage[dvipsnames]{xcolor}
\usepackage{csquotes}
\usepackage[utf8]{inputenc}
\usepackage[english]{babel}
\usepackage{dsfont}
\usepackage{amsthm,amssymb}

\usepackage{bbm}

\usepackage{blkarray}

\usepackage{algorithmicx}
\usepackage{algpseudocode}
\usepackage[norelsize, linesnumbered, ruled, lined, boxed, commentsnumbered]{algorithm2e}
\usepackage{tikz}
\usetikzlibrary{shapes, arrows, positioning,fit,backgrounds,calc}
\usepackage{pgfplots}
\pgfplotsset{compat=1.16}

\usepackage[percent]{overpic}
\usepackage[linesnumbered,ruled]{algorithm2e}
\usepackage[nodisplayskipstretch]{setspace}

\title{}
\author{}
\date{}
\allowdisplaybreaks
\title{
Risk-Aware AoII-Based Scheduling with Hybrid Transmission for a Semi-Markov Source
}
\author{Saeid Sadeghi Vilni, and Risto~Wichman
\thanks{
}
}
\PassOptionsToPackage{silent}{bar}
\begin{document}

\maketitle
\begin{abstract}
We consider a multi-receiver status update system in which a transmitter monitors a finite-state semi-Markov source and decides whether to stay idle, unicast an update, or broadcast a common update. We formulate a risk-aware scheduling problem that minimizes the long-term average sum of the average Age of Incorrect Information (AoII), average risk ratio, and transmission cost. The risk state is defined by whether the AoII exceeds a prescribed threshold. We solve the problem using model-based and model-free policies and compare them with two baselines. Numerical results show that the proposed policies outperform the baselines, exploit both unicast and broadcast transmissions, and capture the effect of the dwell-time law on scheduling performance.
    
\end{abstract}
\vspace{-4mm}
\section{Introduction}
Timely status updates are critical for real-time applications such as industrial automation, intelligent transportation, and smart sensing systems~\cite{kountouris2021semantics}. In these systems, a transmitter observes a physical process and sends status updates
to one or multiple receivers for real-time tracking. The effectiveness of tracking depends not only on the update rate, but also on whether the received information helps the receivers maintain correct estimates
of the source state~\cite{kountouris2021semantics}.

Age of information (AoI) is a widely used metric for quantifying the freshness of received information~\cite{aoi1}.
However, AoI does not directly capture whether the receiver's estimate is correct. To address this limitation, age of incorrect information (AoII) has been introduced as a goal-oriented metric that measures how long the receiver has been operating with an incorrect estimate~\cite{maatouk2020age}. However, in safety- or control-oriented
applications, not all incorrect estimates have the same impact: prolonged incorrectness may expose the system to higher operational
risk~\cite{luo2026exploiting,li2023communication}. This motivates risk-aware status update policies that account not only for the average AoII, but also for the occurrence of high-AoII states, especially in multi-receiver
systems where the transmitter must decide between serving a single receiver and broadcasting a common update to all receivers.

In practical real-time tracking systems, the dwell time of the monitored process in each state can strongly affect the receiver-side estimation accuracy. Standard finite-state Markov models correspond to a specific dwell-time law, whereas semi-Markov models allow more general state-dependent dwell-time distributions~\cite{howard1971dynamic}. This
is useful for physical processes whose operating conditions persist for random durations depending on the system dynamics. Accordingly, in a semi-Markov source model, the future evolution can depend on both the current state and the elapsed time since the last state transition~\cite{qiu2025event}.

Moreover, many real-time monitoring systems involve multiple receivers
tracking the same source~\cite{kadota2018scheduling,
buyukates2019age}. In such systems, the transmitter can either send an
update to a selected receiver with higher reliability or broadcast a
common update to all receivers with lower reliability~\cite{al2022dynamic}.
This creates a scheduling tradeoff between serving the most critical
receiver and reducing the overall risk across the network.

In this work, we consider a multi-receiver status update system
where a transmitter monitors a finite-state semi-Markov source and
sends status updates over error-prone wireless channels. At each slot,
the transmitter decides whether to remain idle, transmit an update to
one selected receiver in unicast mode, or broadcast a common update
to all receivers. Upon successfully decoding an update, each receiver
sends an ACK to the transmitter. The considered system model may represent an intelligent transportation
system where a roadside sensor monitors the traffic state of a road segment, such as free flow, congestion, or incident, and multiple traffic controllers or service units require this information for real-time
operation \cite{al2022dynamic}. Since the duration of each traffic state can vary depending
on traffic demand, weather conditions, and accidents, the source evolution
is naturally captured by a semi-Markov model.

{Status-update systems with multi-user scheduling and hybrid transmission modes have been studied in~\cite{kadota2018scheduling,al2022dynamic}, while semi-Markov source models have been considered for event-triggered communication~\cite{qiu2025event}. In~\cite{kadota2018scheduling}, a base station schedules time-sensitive updates to multiple clients over unreliable broadcast channels, while~\cite{al2022dynamic} considers dynamic unicast-multicast scheduling for vehicular information dissemination. These works are related to the present setting in the use of multi-user wireless update decisions. However, their objectives are AoI-based and, therefore, are driven by update freshness rather than by the source evolution and the correctness of the receiver-side estimates. The work in~\cite{qiu2025event} studied event-triggered communication over AWGN channels for semi-Markov sources. Its main focus is the analysis of a predefined event-triggered policy, including channel coding, error probability, and distortion-power tradeoffs.}

{Motivated by these gaps, we address a risk-aware scheduling problem whose objective is to minimize the long-term average sum of the average AoII, the average risk ratio, and the transmission cost. The risk state of each receiver is defined based on whether its AoII exceeds a prescribed threshold. Unlike~\cite{kadota2018scheduling,al2022dynamic}, the scheduling decisions are driven by the source state, receiver estimates, and elapsed source dwell time, rather than update freshness alone. Unlike~\cite{qiu2025event}, the transmitter dynamically decides whether to remain idle, unicast to one receiver, or broadcast to all receivers, so the semi-Markov source dynamics are embedded in the scheduling state and influence the transmission-mode selection. We model the system as a finite-state average-cost Markov decision process (MDP). We develop a model-based policy using relative value iteration algorithm (RVIA) \cite[Sec. 4.3]{bertks} and a model-free policy using deep Q-network (DQN) \cite{dqn}, and compare them with two baseline policies: a maximum-AoII unicast policy and a threshold-based broadcast policy.}

\section{System Model and Problem Formulation}\label{sysm}
\subsection{System Model}
\begin{figure}
    \centering
    \includegraphics[width=8cm]{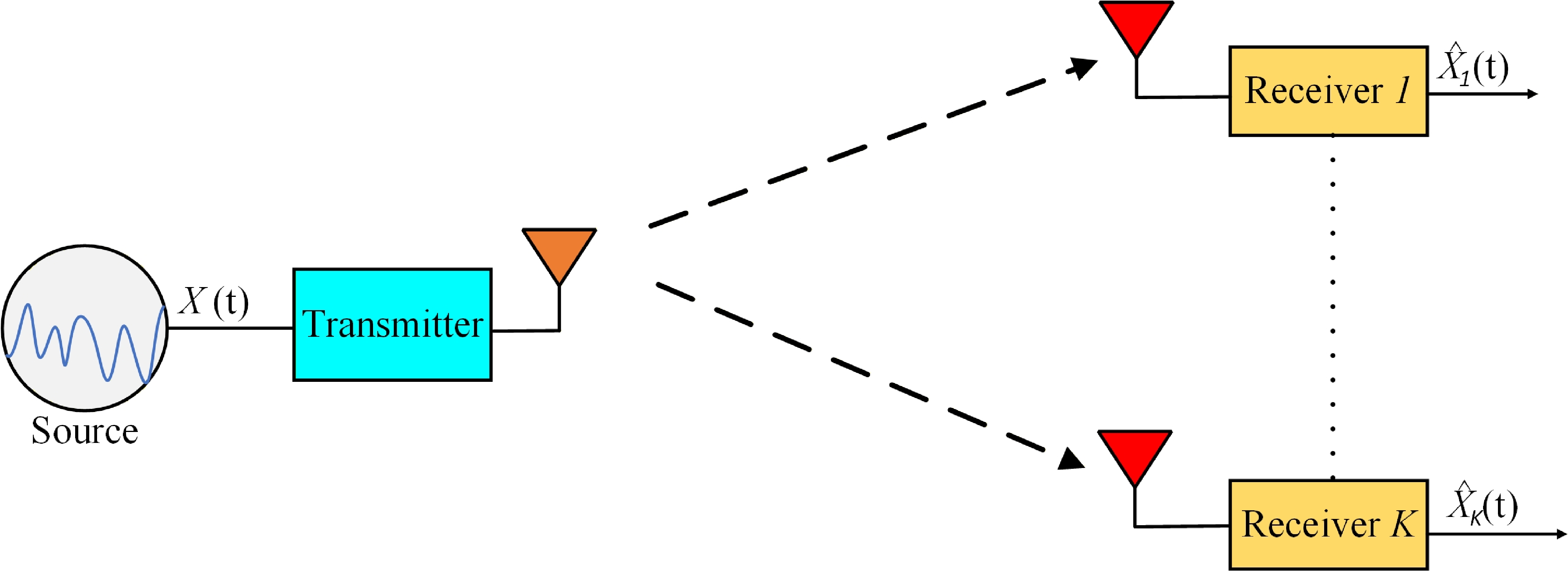}
    \caption{System Model}
    \label{fig:SM}
\end{figure}

We consider a multi-receiver status update system consisting of a transmitter that monitors a physical process and sends updates about the process to $K$ receivers over an error-prone wireless channel, as illustrated in Fig. \ref{fig:SM}. Each receiver is interested in real-time tracking of the source. Time is discrete with unit time slots ${t \in \{1, 2, . . .\}}$. 

The source dynamics are modeled as a finite-state discrete-time semi-Markov process.
Let \(X(t)\in\{1,\ldots,N\}\) denote the source state at slot \(t\).
When the source enters state \(n\), it remains in that state for a random integer-valued dwell time
\(D_n\in\{1,2,\ldots\}\) with (possibly) unbounded support, with probability mass function
\(
f_n(d)\triangleq \Pr(D_n=d), \ d=1,2,\ldots
\).
During the dwell time, the source state remains unchanged.
Upon completion of the dwell time in state \(n\), the source transitions instantaneously
to a next state \(n'\in\{1,\ldots,N\}\) with probability \(p_{nn'}\),
where \(P=[p_{nn'}]\) denotes the embedded Markov chain.
Accordingly, let \(Q_{nn'}(d)\) denote the probability that, after entering state \(n\),
the source remains in \(n\) for exactly \(d\) slots and then transitions to state \(n'\),
which is given by
\begin{align}
&Q_{nn'}(d)
\notag\\&\triangleq
\Pr\!\Big(
X(t+d)=n',\,
X(t+\tau)=n,\ \forall \tau<d
\;\big|\; X(t)=n
\Big)
\notag\\
&=
p_{nn'}\, f_n(d), \qquad d\ge 1,
\end{align}
which fully specifies the source dynamics.

At each slot $t$, the transmitter takes one of the following actions: 1) remain idle, 2) transmit an update to a single receiver (unicast), or 3) transmit an update to all receivers (broadcast). Let $a(t)\in\{0,1,\dots,K,K+1\}$ denote the transmitter action at slot $t$, where $a(t)=0$ indicates that the transmitter stays idle, $a(t)=k$, $1\leq k\leq K$ indicates that the transmitter sends update to $k$th receiver, i.e., unicast mode, and $a(t)=K+1$ indicates that the transmitter sends update to all receivers, i.e., broadcast mode.

The transmitter sends the status update over error-prone wireless
channels. We assume identical packet success probabilities, with
independent receptions, across receivers. In unicast mode, the update
is intended for a selected receiver and is transmitted with a higher
packet success probability, since the transmission can be optimized
for that receiver, e.g., through beamforming.
Nevertheless, due to the broadcast nature of the wireless medium, the
non-intended receivers may also overhear and decode the update, but
only with a much smaller probability. In broadcast mode, the same
update is intended for all receivers and must be decodable by multiple
receivers, which results in a lower packet success probability compared
with unicast. Let $p_{\mathrm{s}}^{\mathrm{uni}}$ denote the packet
success probability of the intended receiver in unicast mode, let
$p_{\mathrm{s}}^{\mathrm{oh}}$ denote the overhearing success
probability of a non-intended receiver in unicast mode, and let
$p_{\mathrm{s}}^{\mathrm{bc}}$ denote the packet success probability
in broadcast mode, where
$p_{\mathrm{s}}^{\mathrm{uni}}>p_{\mathrm{s}}^{\mathrm{bc}}>
p_{\mathrm{s}}^{\mathrm{oh}}$. After each transmission, each receiver
that successfully decodes the update sends an acknowledgment (ACK) to
the transmitter; the feedback channels are assumed to be instantaneous
and error-free. Let $d_k(t)\in\{0,1\}$ denote the packet reception
status at receiver $k$ in slot $t$, where $d_k(t)=1$ indicates that
receiver $k$ successfully decodes the update, either as the intended
receiver, through overhearing, or through broadcast, and $d_k(t)=0$
otherwise.

Each receiver estimates the source state based on the status updates
received from the transmitter. Since the source is semi-Markov, the
future source evolution depends not only on the current source state,
but also on how long the source has already remained in that state.
Therefore, we introduce $\delta_{\mathrm{x}}(t)$ as the elapsed time
since the most recent state transition of the source at slot $t$. Its
evolution is given by
\begin{align}
\delta_{\mathrm{x}}(t+1) &=
\begin{cases}
0, & X(t+1)\neq X(t),\\
\min\{\delta_{\mathrm{x}}(t)+1,\delta_{\mathrm{x}}^{\max}\},
& X(t+1)=X(t),
\end{cases}
\label{eq:source_age_evol}
\end{align}
where $\delta_{\mathrm{x}}^{\max}$ is the upper bound of
$\delta_{\mathrm{x}}(t)$. We adopt the common assumption (see, e.g.,
\cite{gz,mmpower,ssvtvt}) that age-related variables are
upper-bounded by finite constants. This assumption ensures a finite state
space and is justified by the fact that sufficiently large age values
provide no additional timeliness information for decision-making.

We assume that each successfully decoded status update contains the
current augmented source state $(X(t),\delta_{\mathrm{x}}(t))$. Let
$Y_k(t)=(Y_k^{\mathrm{x}}(t),Y_k^{\mathrm{m}}(t))$ denote the most
recently decoded augmented source state at receiver $k$, and let
$\delta_k(t)$ denote the elapsed time since this update was decoded.
Since ACK feedback is available, the transmitter can track
$Y_k(t)$ and $\delta_k(t)$ for each receiver. These variables evolve as
\begin{align}
Y_k(t+1) &=
\begin{cases}
(X(t),\delta_{\mathrm{x}}(t)), & d_k(t)=1,\\
Y_k(t), & d_k(t)=0,
\end{cases}
\label{eq:Yk_evol}
\\
\delta_k(t+1) &=
\begin{cases}
1, & d_k(t)=1,\\
\min\{\delta_k(t)+1,\delta^{\max}\}, & d_k(t)=0.
\end{cases}
\label{eq:receiver_age_evol}
\end{align}
Here, $\delta_k(t)$ is the age of the most recently decoded augmented
source state at receiver $k$; it determines how many prediction steps
are needed to infer the receiver's current belief.

Let $b_{k,n,m}(t)$ denote the probability assigned by receiver $k$ to
the event $\{X(t)=n,\delta_{\mathrm{x}}(t)=m\}$. Since
$Y_k(t)$ and $\delta_k(t)$ fully determine the receiver's information,
we write
\[
b_{k,n,m}(t)
\triangleq
\Pr\big(X(t)=n,\delta_{\mathrm{x}}(t)=m
\mid Y_k(t),\delta_k(t)\big).
\]
The marginal probability of source state $n$ is
\[
\pi_{k,n}(t)=
\sum_{m=0}^{\delta_{\mathrm{x}}^{\max}} b_{k,n,m}(t),
\]
and receiver $k$ employs the maximum-likelihood (ML) estimator
\[
\hat X_k(t)=
\arg\max_{n\in\{1,\ldots,N\}} \pi_{k,n}(t).
\]

To compute the belief induced by $Y_k(t)$ and $\delta_k(t)$, define
\[
\bar F_n(m)\triangleq \Pr(D_n>m)
=
\sum_{\ell=m+1}^{\infty} f_n(\ell).
\]
If the source is currently in state $n$ and has already stayed in this
state for $m$ slots, then the event $D_n>m$ is known. Hence, the
probability that the source remains in state $n$ for one more slot is
\[
\Pr(D_n>m+1\mid D_n>m)
=
\frac{\bar F_n(m+1)}{\bar F_n(m)}.
\]
Similarly, the probability that the source leaves state $n$ in the next
slot and moves to state $n'\neq n$ is
\[
p_{nn'}\Pr(D_n=m+1\mid D_n>m)
=
p_{nn'}\frac{f_n(m+1)}{\bar F_n(m)}.
\]
Equivalently, using $Q_{nn'}(d)=p_{nn'}f_n(d)$, for
$i,n\in\{1,\ldots,N\}$ and
$j,m\in\{0,\ldots,\delta_{\mathrm{x}}^{\max}\}$, the one-step
transition probability of the augmented source state is
\begin{align}
&\Pr\big((i,j)\rightarrow(n,m)\big)=\notag\\&
\begin{cases}
\dfrac{\bar F_i(j+1)}{\bar F_i(j)},
& n=i,\; m=\min\{j+1,\delta_{\mathrm{x}}^{\max}\},\\[2ex]
\dfrac{Q_{in}(j+1)}{\bar F_i(j)},
& n\neq i,\; m=0,\\[2ex]
0, & \text{otherwise}.
\end{cases}
\label{eq:aug_transition}
\end{align}

Accordingly, the belief $b_{k,n,m}(t)$ can be obtained by initializing
the distribution at the last decoded augmented source state $Y_k(t)$
and applying the transition rule in \eqref{eq:aug_transition} for
$\delta_k(t)$ prediction steps. Thus, the full belief vector is not
stored as an independent state variable; instead, it is reconstructed
from $(Y_k(t),\delta_k(t))$ whenever the ML estimate is computed.

\subsection{Problem Formulation}
Our objective is to minimize the long-term average cost comprising the Age of Incorrect Information (AoII) at the receivers, a risk-aware cost, and the transmission cost associated with unicast and broadcast updates, by optimizing the transmitter’s scheduling decisions. At each slot $t$, the transmitter selects an action based on the
available information, including the current augmented source state and
the receiver-side information states tracked through ACK feedback.

Let \(\Delta_k(t)\) denote the AoII at receiver \(k\) at slot \(t\), defined as the elapsed time since the most recent slot at which the receiver’s estimate coincided
with the true source state, i.e., $
\Delta_k(t) = t - \ell_k(t),
\quad
\ell_k(t) \triangleq \max\{t' \le t : X(t') = \hat X_k(t')\}
$. To simplify notation, let us denote
\(
\tilde{\Delta}_k(t) \triangleq \min\{\Delta_k(t)+1, \delta^{\max}\}.
\)
The AoII evolution at receiver \(k\) is then given by
\begin{align}
&\Delta_k(t\!+\!1)\!=\!\notag\\&
\begin{cases}
0, 
& d_k(t)\!=\!1,\; X(t\!+\!1)\!=\!X(t), \\[2pt]
\tilde{\Delta}_k(t), 
& d_k(t)\!=\!1,\; X(t\!+\!1)\!\neq\! X(t), \\[2pt]
\mathbbm{1}_{\{\Delta_k(t)\neq 0\}}\tilde{\Delta}_k(t),
& d_k(t)\!=\!0,\; X(t\!+\!1)\!=\!X(t), \\[2pt]
\mathbbm{1}_{\{X(t\!+\!1)\neq \hat X_k(t)\}}\tilde{\Delta}_k(t),
& d_k(t)\!=\!0,\; X(t\!+\!1)\!\neq\! X(t).
\end{cases}
\end{align}
We evaluate performance in terms of the average AoII across receivers, given as $\bar{\Delta}(t) \triangleq \frac{1}{K}\sum_{k=1}^{K} \Delta_k(t)$.

To explicitly capture the risk state, we introduce a threshold-based risk indicator.
A receiver is considered at a \emph{risk state} whenever its AoII exceeds a prescribed
threshold \(\Delta^{\mathrm{th}}\). Let $r_k(t)$ denote the risk indicator of receiver $k$ at slot $t$, given as
\begin{equation}
r_k(t) \triangleq \mathbbm{1}_{\{\Delta_k(t) > \Delta^{\mathrm{th}}\}}.
\end{equation}
The average risk across receivers defined as $\bar{r}(t) \triangleq \frac{1}{K}\sum_{k=1}^{K} r_k(t)$.

Let $c_{\rm tx}$ denote the cost of one transmission attempt. The
per-slot transmission cost is
\begin{equation}
c(t)=
\begin{cases}
0, & a(t)=0,\\
c_{\rm tx}, & a(t)\in\{1,\ldots,K+1\}.
\end{cases}
\end{equation}
Thus, the cost penalizes channel usage, while the difference between
unicast and broadcast is reflected through the reception probabilities.

Let $g(t)=\bar{\Delta}(t)+\gamma \bar r(t)+\lambda c(t)$ denot the objective cost at slot $t$, the considered problem can be formulated as the following stochastic control problem:
\begin{subequations}\label{pmain}
\begin{alignat}{2}
\displaystyle\amin_{} \quad & \underset{T\rightarrow \infty}{\limisup} ~~~ \frac{1}{T} \sum_{t=1}^{T}\mathbb{E} \{ g(t)\}\\
\mbox{subject to} \quad &a(t)\in \{0,1,\dots,K, K+1\}\label{p1:2},
\end{alignat}
\end{subequations}
with variables $a(t)$ for all $t\in\{1,2, \dots\}$. The coefficients $\gamma$ and $\lambda$ are design constants. The expression \eqref{p1:2} indicates the possible values of the decision variable. In problem \eqref{pmain}, $\mathbb{E}\{\cdot\}$ is the expectation with respect to the randomness of the system (i.e., randomness in the source dynamic and the communication channels) and the decision variable $a(t)$.

\section{Transmission Scheduling Policy}

To solve problem~\eqref{pmain}, we formulate the system as a finite-state
average-cost MDP. The state at slot $t$ is defined as
\begin{align}
s(t)=\big(X(t),\delta_x(t),\{Y_k(t),\delta_k(t),\Delta_k(t)\}_{k=1}^{K}\big),
\end{align}
where the bounded age variables make the state space finite. The state
contains the current augmented source state, the receiver-side decoded
information tracked through ACK feedback, and the AoII values. The ML
estimate $\hat X_k(t)$ is not included as an independent state variable,
since it is reconstructed from $(Y_k(t),\delta_k(t))$ using the transition
rule in~\eqref{eq:aug_transition}. The action space is
$\mathcal{A}=\{0,1,\ldots,K,K+1\}$, and the one-stage cost is $g(t)$.


When the source and channel statistics are known, we apply relative value iteration algorithm \cite[Sec. 4.3]{bertks} over the truncated finite state space and obtain a stationary scheduling policy, referred to as the stationary scheduling policy (SSP). The policy maps each state $s(t)$ to an action in $\mathcal{A}$.

For unknown dynamics, we utilize DQN~\cite{dqn} to derive a learning-based scheduling policy (LSP). The DQN takes $s(t)$ as input and estimates the action-value function for all actions in $\mathcal{A}$, with the reward defined as the negative of the one-stage cost, i.e., $-g(s(t),a(t))$. The implementation details are provided in Section~IV.

As benchmarks, we consider two heuristic policies. The
maximum-AoII unicast policy (MAUP) always serves the receiver with the
largest AoII, i.e.,
\[
a(t)=\arg\max_{k\in\{1,\ldots,K\}}\Delta_k(t).
\]
The threshold-based broadcast policy (TBP) broadcasts when the number of
receivers in the risk state exceeds a threshold $m$; otherwise, it serves
the receiver with the largest AoII:
\[
a(t)=
\begin{cases}
K+1, & \sum_{k=1}^{K}\mathbbm{1}_{\{\Delta_k(t)>\Delta^{\rm th}\}}\ge m,\\
\arg\max_{k}\Delta_k(t), & \text{otherwise}.
\end{cases}
\]
These benchmarks are used to evaluate the benefit of adaptive
AoII- and risk-aware hybrid transmission scheduling.

\section{Numerical Results}\label{snrs}
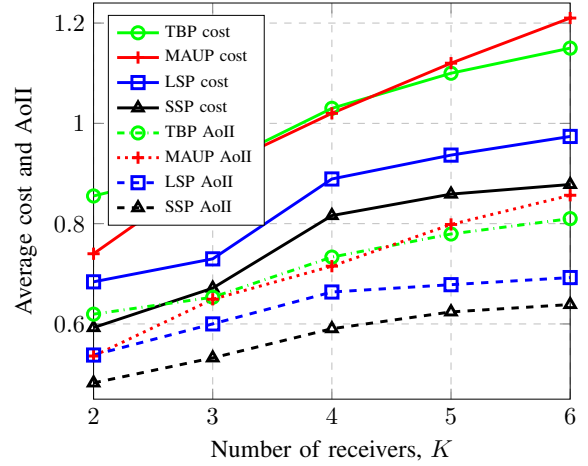
\begin{figure}[!t]
\centering
\begin{tikzpicture}[scale=0.92]
\begin{axis}[
    xlabel={Number of receivers, $K$},
    ylabel={Average cost and AoII},
    xmin=2, xmax=6,
    ymin=0.45, ymax=1.24,
    xtick={2,3,4,5,6},
    grid=both, grid style=dashed,
    legend style={at={(0.03,0.97)},anchor=north west,font=\scriptsize},
    legend cell align=left,
]


\addplot+[
color=green,solid,
mark=o,mark options={solid},
line width=1.2pt,
mark size=2.5pt,
]
coordinates {
(2,0.8552)(3,0.9129)(4,1.03)(5,1.1)(6,1.15)
};

\addplot+[
color=red,solid,
mark=+,mark options={solid},
line width=1.2pt,
mark size=2.5pt,
]
coordinates {
(2,0.74)(3,0.9073)(4,1.02)(5,1.12)(6,1.21)
};

\addplot+[
color=blue,solid,
mark=square,mark options={solid},
line width=1.2pt,
mark size=2.5pt,
]
coordinates {
(2,0.684)(3,0.7298)(4,0.8891)(5,0.9367)(6,0.9739)
};

\addplot+[
color=black,solid,
mark=triangle,mark options={solid},
line width=1.2pt,
mark size=2.5pt,
]
coordinates {
(2,0.5927)(3,0.6714)(4,0.8161)(5,0.8589)(6,0.8781)
};


\addplot+[
color=green,dashdotted,
mark=o,mark options={solid},
line width=1.2pt,
mark size=2.5pt,
opacity=0.55,
]
coordinates {
(2,0.6195)(3,0.6533)(4,0.7333)(5,0.7794)(6,0.81)
};

\addplot+[
color=red,dotted,
mark=+,mark options={solid},
line width=1.2pt,
mark size=2.5pt,
opacity=0.55,
]
coordinates {
(2,0.536)(3,0.6492)(4,0.7152)(5,0.7984)(6,0.8566)
};

\addplot+[
color=blue,dashed,
mark=square,mark options={solid},
line width=1.2pt,
mark size=2.5pt,
]
coordinates {
(2,0.538)(3,0.6001)(4,0.6637)(5,0.6783)(6,0.6925)
};

\addplot+[
color=black,dashed,
mark=triangle,mark options={solid},
line width=1.2pt,
mark size=2.5pt,
]
coordinates {
(2,0.4824)(3,0.5323)(4,0.5905)(5,0.6240)(6,0.6387)
};

\legend{
TBP cost,
MAUP cost,
LSP cost,
SSP cost,
TBP AoII,
MAUP AoII,
LSP AoII,
SSP AoII
}
\end{axis}
\end{tikzpicture}
\caption{Average cost and average AoII versus the number of receivers $K$ for different policies, where $N=2$, $\delta_{\mathrm{x}}^{\max}=8$, $\delta_k^{\max}=8$, $\Delta^{\max}=8$, $\rho=0.25$, $p_{\mathrm{s}}^{\mathrm{uni}}=0.85$, $p_{\mathrm{s}}^{\mathrm{oh}}=0.08$, $p_{\mathrm{s}}^{\mathrm{bc}}=0.45$, $c_{\mathrm{tx}}=1$, $\gamma=2$, $\lambda=0.1$, and $\Delta^{\mathrm{th}}=2$.}
\label{fig:cost_aoii_perK}
\end{figure}

In this section, we evaluate the system under different policies and settings. Unless otherwise stated, the dwell time in state $n$ follows $f(d)=\rho(1-\rho)^{d-1}$, $d\geq 1$, where $1/\rho_n$ is the mean dwell time. For DQN, the discount factor is $0.95$, the replay-buffer size is $5\times 10^4$, the mini-batch size is $64$, and learning starts after $1000$ samples. Training is performed every four slots with learning rate $10^{-3}$, and the target network is updated every $500$ slots. The exploration probability decreases from $1$ to $0.05$ over $15000$ slots. 
Fig.~\ref{fig:cost_aoii_perK} shows the average cost and average AoII versus the number of receivers $K$. Both metrics increase with $K$, since the transmitter must maintain accurate estimates at more receivers under the same transmission opportunities. The SSP achieves the best performance, as it uses the known source and channel statistics to jointly select between idle, unicast, and broadcast actions. The LSP follows the same trend and consistently outperforms the MAUP and TBP benchmarks, showing that the learned policy captures the main scheduling tradeoffs without explicit knowledge of the transition model. The gap between the proposed policies and the benchmarks confirms the benefit of adaptive AoII-aware hybrid transmission scheduling.

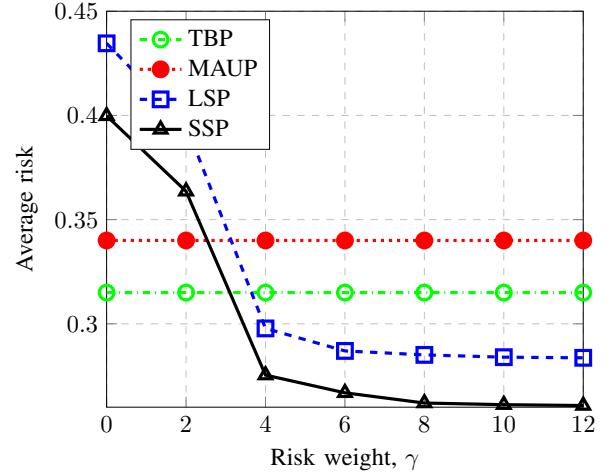
\begin{figure}[!t]
\centering
\begin{tikzpicture}[scale=0.92]
\begin{axis}[
    xlabel={Risk weight, $\gamma$},
    ylabel={Average risk},
    xmin=0, xmax=12,
    ymin=0.26, ymax=0.45,
    xtick={0,2,4,6,8,10,12},
    grid=both, grid style=dashed,
    legend style={at={(0.05,0.65)},anchor=south west},
    legend cell align=left,
]

\addplot+[
color=green,dashdotted,
mark=o,mark options={solid},
line width=1.3pt,
mark size=3pt,
]
coordinates {
(0,0.315)(2,0.315)(4,0.315)(6,0.315)(8,0.315)(10,0.315)(12,0.315)
 };

\addplot+[
color=red,dotted,
mark=*,mark options={solid},
line width=1.3pt,
mark size=3pt,
]
coordinates {
(0,0.34)(2,0.34)(4,0.34)(6,0.34)(8,0.34)(10,0.34)(12,0.34)
 };

\addplot+[
color=blue,dashed,
mark=square,mark options={solid},
line width=1.3pt,
mark size=3pt,
]
coordinates {
(0,0.4346)(2,0.3953)(4,0.2978)(6,0.2870)(8,0.2851)(10,0.284)
(12,0.2837)
 };

\addplot+[
color=black,
mark=triangle,mark options={solid},
line width=1.3pt,
mark size=3pt,
]
coordinates {
(0,0.3998)(2,0.3636)(4,0.2754)(6,0.2669)(8,0.2620)(10,0.2612)
(12,0.2608)
 };

\legend{TBP,MAUP,{LSP},SSP}
\end{axis}
\end{tikzpicture}
\caption{Average risk ratio versus the risk-weight parameter $\gamma$
for different policies, where $K=8$, $N=4$,
$\delta_{\mathrm{x}}^{\max}=8$, $\delta_k^{\max}=8$,
$\Delta^{\max}=8$, $\rho=0.5$,
$p_{\mathrm{s}}^{\mathrm{uni}}=0.5$,
$p_{\mathrm{s}}^{\mathrm{oh}}=0.05$,
$p_{\mathrm{s}}^{\mathrm{bc}}=0.25$,
$c_{\mathrm{tx}}=1$, $\lambda=1.5$, and
$\Delta^{\mathrm{th}}=3$.}
\label{fig:perGperP}
\end{figure}

Fig.~\ref{fig:perGperP} shows the impact of the risk-weight parameter $\gamma$ on the average risk ratio. The TBP and MAUP policies transmit at every slot and are not directly optimized with respect to $\gamma$; therefore, their risk ratios remain almost unchanged. In contrast, the SSP and LSP policies adapt their scheduling decisions as $\gamma$ increases. Since a larger $\gamma$ assigns a higher penalty to receivers whose AoII exceeds the threshold, both policies reduce the average risk ratio by prioritizing risky receivers and selecting the transmission mode accordingly. The SSP achieves the lowest risk ratio for all values of $\gamma$, while the LSP follows the same trend and approaches the SSP performance for large $\gamma$. This confirms that the proposed risk-aware formulation enables the scheduler to control the occurrence of high-AoII states.

\begin{figure}[t]
\centering
\begin{tikzpicture}[scale=0.92]
\begin{axis}[
    ybar stacked,
    width=0.95\columnwidth,
    height=0.62\columnwidth,
    bar width=16pt,
    ymin=0,
    ymax=1,
    ylabel={Action-selection ratio},
    xlabel={Number of receivers $K$},
    symbolic x coords={$2$,$4$,$6$,$8$},
    xtick=data,
    grid=both,
    legend style={
        at={(0.5,1.03)},
        anchor=south,
        legend columns=3,
        font=\footnotesize
    },
    tick label style={font=\footnotesize},
    label style={font=\footnotesize},
]
\addplot coordinates {($2$,0.253) ($4$,0.1809) ($6$,0.010) ($8$,0.055)};
\addplot coordinates {($2$,0.275) ($4$,0.6248) ($6$,0.7487) ($8$,0.039)};
\addplot coordinates {($2$,0.472) ($4$,0.1941) ($6$,0.2406) ($8$,0.906)};
\legend{Idle, Unicast, Broadcast}
\end{axis}
\end{tikzpicture}
\caption{Action-selection ratios versus number of receivers $K$, where
$N=4$, $\delta_{\mathrm{x}}^{\max}=8$, $\delta_k^{\max}=8$,
$\Delta^{\max}=8$, $\rho=0.5$,
$p_{\mathrm{s}}^{\mathrm{uni}}=0.6$,
$p_{\mathrm{s}}^{\mathrm{oh}}=0.05$,
$p_{\mathrm{s}}^{\mathrm{bc}}=0.35$,
$c_{\mathrm{tx}}=1$, $\gamma=6$, $\lambda=1$, and
$\Delta^{\mathrm{th}}=4$.}
\label{fig:action_ratio_K}
\end{figure}
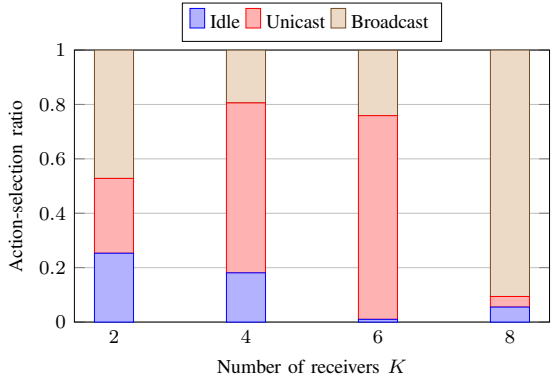

Fig.~\ref{fig:action_ratio_K} illustrates the action-selection behavior of the LSP for different numbers of receivers. Although the
figure is obtained from the learned DQN policy, it reflects the scheduling behavior enabled by the proposed hybrid action space. As $K$ increases, the idle ratio generally decreases, showing that the scheduler
uses the channel more aggressively when more receivers track the source.
The unicast-broadcast split depends on the reliability tradeoff between
the two modes: unicast is useful for serving a critical receiver with
higher success probability, whereas broadcast becomes attractive when a
common update can benefit several receivers. These results show that the
learned policy exploits both transmission modes instead of relying on a
single fixed action.

To examine the impact of the dwell-time law, we compare three choices of
$f(d)$ with the same mean dwell time. The first one is
$f^{(1)}(d)=\rho(1-\rho)^{d-1}$, $d\geq 1$, where $1/\rho$ is the
mean dwell time. The second one is the deterministic dwell time
$f^{(2)}(d)=1$ for $d=\bar D$ and $f^{(2)}(d)=0$ otherwise. The third
one is the truncated power-law dwell time
$f^{(3)}(d)=d^{-\alpha}/\sum_{\ell=1}^{D_{\max}}\ell^{-\alpha}$,
$d=1,\ldots,D_{\max}$, with $f^{(3)}(d)=0$ for $d>D_{\max}$. For
$f^{(3)}(d)$, $\alpha$ is selected for each target mean dwell time such
that $\sum_{d=1}^{D_{\max}} d f^{(3)}(d)$ matches the desired mean.

\begin{figure}[!t]
\centering
\begin{tikzpicture}[scale=0.92]
\begin{axis}[
    xlabel={Mean dwell time, $\mathbb{E}[D_n]$},
    ylabel={Average cost},
    xmin=2, xmax=10,
    ymin=0.3, ymax=3.95,
    xtick={2,4,6,8,10},
    grid=both, grid style=dashed,
    legend style={at={(0.98,0.98)},anchor=north east},
    legend cell align=left,
]

\addplot+[
color=blue,dashed,
mark=square,mark options={solid},
line width=1.5pt,
mark size=3pt,
]
coordinates {
(2,1.3790)(4,0.9221)(6,0.7095)(8,0.5258)(10,0.4464)
};

\addplot+[
color=red,dotted,
mark=*,mark options={solid},
line width=1.5pt,
mark size=3pt,
]
coordinates {
(2,3.8282)(4,1.4468)(6,0.8788)(8,0.6012)(10,0.4821)
};

\addplot+[
color=black,
mark=triangle,mark options={solid},
line width=1.5pt,
mark size=3pt,
]
coordinates {
(2,0.8144)(4,0.7310)(6,0.5892)(8,0.4440)(10,0.3721)
};

\legend{
$f^{(1)}(d)$,
$f^{(2)}(d)$,
$f^{(3)}(d)$
}
\end{axis}
\end{tikzpicture}
\caption{Average cost versus the mean dwell time for different dwell-time
laws under the LSP, where $K=4$, $N=2$, $\delta_{\mathrm{x}}^{\max}=15$,
$\delta_k^{\max}=15$, $\Delta^{\max}=15$, $p_{\mathrm{s}}^{\mathrm{uni}}=0.85$,
$p_{\mathrm{s}}^{\mathrm{oh}}=0.08$, $p_{\mathrm{s}}^{\mathrm{bc}}=0.45$,
$c_{\mathrm{tx}}=1$, $\gamma=2$, $\lambda=0.1$, $\Delta^{\mathrm{th}}=2$,
and $D_{\max}=100$ for truncated power-law dwell time.}
\label{fig:dwell_law_effect}
\end{figure}
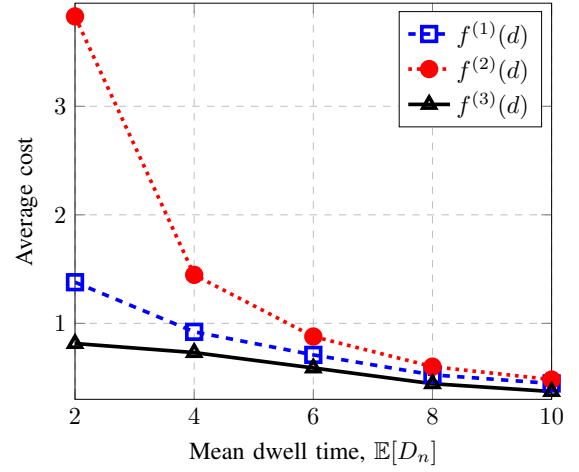

Fig.~\ref{fig:dwell_law_effect} shows the average cost versus the mean
dwell time for the three dwell-time laws. As the mean dwell time
increases, the average cost decreases for all cases, since the source
changes less frequently and the receivers can maintain more accurate
estimates. The curves are different even under the same mean dwell time,
which shows that the dwell-time law itself affects the scheduling
performance. In particular, $f^{(2)}(d)$ yields the largest cost for
small mean dwell times, while $f^{(3)}(d)$ provides the lowest cost over
most of the considered range. This confirms that the proposed formulation
captures both the mean dwell time and the underlying dwell-time law.

\vspace{-0.5cm}  
\section{Conclusions}\label{clc}
We considered a risk-aware multi-receiver status update system with a finite-state semi-Markov source and hybrid unicast/broadcast transmissions. We formulated the scheduling problem as the minimization of the long-term average sum of the average AoII, the average risk ratio, and the transmission cost. We solved the problem using model-based and model-free scheduling policies and compared them with two fixed baseline policies. 
The results also showed that increasing the risk-weight parameter reduces the occurrence of high-AoII states, and that the learned policy uses both unicast and broadcast decisions depending on the number of receivers and the risk distribution. Furthermore, the results demonstrated that the dwell-time law affects the scheduling performance even when the mean dwell time is fixed, highlighting the importance of the semi-Markov source model.

\vspace{-4mm}
\bibliographystyle{IEEEtran}
\bibliography{short-conf,short-jour,Main}

\end{document}